# Unplug, Mute, Avoid: Investigating smart speaker users' privacy protection behaviours in Saudi Homes

**Abdulrhman Alorini,** School of Computer Science, University of Technology Sydney, Sydney, NSW, Australia

**Yufeng Wu,** School of Computer Science, University of Technology Sydney, Sydney, NSW, Australia

**Abdullah Bin Sawad,** Department of CIT, The Applied College - King Abdullaziz University, Jeddah, Saudi Arabia

**Mukesh Prasad,** Computer Science**,** University of Technology Sydney, Sydney, NSW, Australia

**A. Baki Kocaballi,** School of Computer Science, University of Technology Sydney, Sydney, Australia

Smart speakers are increasingly integrated into domestic life worldwide, yet their privacy risks remain underexplored in non-Western cultural contexts. This study investigates how Saudi Arabian users of smart speakers navigate privacy concerns within collectivist, gendered, and often multigenerational households. Using cultural probes followed by semi-structured interviews with 16 participants, we uncover everyday privacy-protective behaviours including unplugging devices, muting microphones, and avoiding voice interactions altogether. These practices are shaped not only by individual risk perceptions but also by household norms, room configurations, and interpersonal dynamics. We contribute empirical insights from an underrepresented region, theoretical extensions to contextual integrity frameworks, and design directions for culturally responsive voice interfaces. This work expands the global conversation on smart speaker privacy and informs more inclusive HCI practices in increasingly diverse smart home environments.

## 1 INTRODUCTION

The global adoption of smart speakers has risen significantly, driven by the convenience of using natural language that enable users to interact with devices seamlessly and manage other connected devices remotely [1]. These smart speakers, however, have capabilities for gathering and storing data, such as audio and, in some cases, video, from their surroundings,

---

Authors' Contact Information: Abdulrhman Alorini, abdulrhmanfahada.f.alorini@student.uts.edu.au, School of Computer Science, University of Technology Sydney, Sydney, NSW, Australia; Yufeng Wu, School of Computer Science, University of Technology Sydney, Sydney, NSW, Australia; Abdullah Bin Sawad, Department of CIT, The Applied College - King Abdullaziz University, Jeddah, Saudi Arabia, asawad@kau.edu.sa; Mukesh Prasad, School of Computer Science, University of Technology Sydney, Sydney, NSW, Australia; A. Baki Kocaballi, Baki.Kocaballi@uts.edu.au, School of Computer Science, University of Technology Sydney, Sydney, NSW, Australia which raises serious privacy and security concerns. Given that homes are traditionally private spaces, the boundaries of users' privacy rights concerning smart speaker usage remain ambiguous. Additionally, with technology advancing rapidly, social and cultural norms around privacy are in a state of constant evolution, making it challenging to establish consistent and widely accepted privacy practices in this area [2].

Most research on smart speaker privacy has been conducted in Western regions, primarily in the United States and Europe [3, 4]. The results of these studies reveal that users often have to choose between convenience and privacy. A concern arises in shared or multi-user households due to constant monitoring, unclear data practices, as well as an inability to control who has access to or uses the device. Data collection and sharing are often not understood by users, both owners and visitors, resulting in tensions and informal social norms around the use of devices [4]. While smart speakers are widely appreciated by older adults due to their abilities to support their health, memory, and accessibility, concerns persist regarding eavesdropping and the misuse of information [3]. In general, there is a desire for greater transparency and greater control over data and privacy settings across user groups. However, despite the widespread adoption of smart speakers in non-Western settings, such as Arab countries, there remains a significant lack of studies in these contexts, limiting our understanding of how unique cultural frameworks shape user behaviour and perceptions. This paper addresses this critical gap by presenting findings from a qualitative study examining the privacy protection behaviours of smart speaker users in the Kingdom of Saudi Arabia. This context is particularly insightful because Saudi Arabia is a society deeply shaped by unique social norms and customs rooted in its religious and cultural heritage [5]. Studying this demographic offers more than just filling a geographical void; it provides an opportunity to understand how deep-rooted cultural and religious tenets such as those pertaining to family, modesty, and community fundamentally shape user interactions with pervasive technologies like smart speakers and their associated privacy perceptions. These insights have the potential to challenge and enrich existing HCI models of privacy, which often lack the nuanced perspectives derived from such distinctly collectivist and faith-oriented societies.

To address the underexplored privacy experiences of smart speaker users in Saudi Arabia, we conducted a qualitative study involving 16 participants, using cultural probes and semi-structured interviews to surface everyday privacy protection behaviours. Thematic analysis of the data revealed five key themes, including speaker placement, usage patterns, specific privacy concerns, mitigation strategies, and responses to tangible privacy solutions. Our findings inform culturally grounded design recommendations for privacy-preserving smart speaker interactions in Saudi households and similar socio-cultural settings.

## 2 BACKGROUND

### 2.1 Privacy

The concept of privacy has evolved rapidly in the digital age, becoming more complex and multifaceted as a result. Defining privacy is challenging due to its many dimensions and contexts. Historically, Weston's definition of privacy is regarded as one of the earliest definitions of privacy, as it states: The right of the individual or institution to determine how, to whom, and to what extent personal information is shared [6]. A technical definition of this would be that it is an information security service that protects personal information, such as preferences and characteristics, associated with individuals' identities from unauthorized access or disclosure [7]. A comprehensive definition can be constructed in the following way based on the definitions provided by [6, 7]: an individual's right to control access to his or her personal information, spaces, and experiences, free from unauthorized intrusion or surveillance, as well as the ability to maintain personal boundaries, maintain confidentiality and anonymity as desired, and maintain their independence and dignity at all times.

Several dimensions of privacy are discussed in the previous definition, including: informational privacy, which describes how personal information is collected, used, and disseminated [8]. Physique privacy refers to protecting one's physical space and body against intrusion [9]. A person's social privacy includes managing their social relationships and interactions, which includes the ability to withdraw from social situations when necessary [10]. A psychological privacy is the ability to maintain control over cognitive and affective processes, as well as to protect the self from external intrusion [9].

In recent years, there has been a growing connection between technology development and privacy. This has become increasingly evident with the proliferation of Internet of Things (IoT) devices that collect and analyse data in various settings. In this context, the theory of privacy as contextual integrity [11] emphasizes that privacy expectations are shaped by the specific context in which data is collected and used. Specifically, it consists of two criteria: appropriate data flow and appropriate data collection. Whenever either of these criteria is violated, a privacy breach has occurred. As a result, this theory has become increasingly important in the context of smart homes and the Internet of Things. As a result, contextual integrity suggests that privacy expectations are context-specific and are influenced by social norms and practices [8]. The privacy calculus theory is one of several theoretical perspectives that contribute to our understanding of privacy. According to this theory, individuals weigh their perceived risks against the potential benefits of disclosing personal information [10]. As another theory, institutional logic theory examines the ways in which institutional structures and practices influence privacy perceptions and behaviours [12]. It is also important to take into account the cultural dimensions theory, which recognizes that privacy expectations and practices may differ significantly from one culture to another [12].

In summary, this subsection provides an overview of how privacy is defined and theorized across different disciplines, highlighting its multifaceted nature and contextual sensitivity. These perspectives particularly contextual integrity, privacy calculus, and cultural dimensions offer a theoretical foundation for understanding users' privacy behaviours and expectations, which are central to this study's focus on smart speaker usage in culturally specific contexts.

### 2.2 Privacy in smart speakers

Smart speakers (e.g., Amazon Echo, Google Nest, Apple HomePod) are AI-powered, voice-activated devices capable of performing a range of tasks such as playing media, managing schedules, controlling smart home systems, and facilitating online services. Their always-on microphones, cloud-based voice recognition, and machine learning capabilities enhance responsiveness but raise significant privacy concerns—especially due to their integration into intimate domestic spaces [13]. Voice data is particularly sensitive, as it can reveal not only content but also emotional state and health information [14]. The broader societal and long-term privacy implications of widespread smart speaker use remain uncertain [15], and users often lack confidence in the effectiveness of available privacy controls [16]. Studies by Lau [17] and Meng [4] highlight bystander privacy, device placement, and informed consent as critical issues, especially in shared spaces.

Several general principles may apply to smart speakers, such as contextual integrity, which suggests that privacy expectations vary depending on the user's location, activities, and cultural norms [18]. Furthermore, it is important to take privacy by design into account by integrating privacy-enhancing features directly into smart speaker hardware and software [19]. Additionally, providing intuitive, accessible privacy controls is essential, but due to the lack of traditional interfaces, this may prove challenging. Transparency needs to be provided by providing clear communication concerning what data is collected and how it is used. However, the model of voice-based interaction makes this task more challenging.

### 2.3 The concept of privacy in the Global South

Privacy concerns related to smart speakers in the Global South remain understudied, despite increasing adoption in developing countries. Users in these regions face unique vulnerabilities due to limited digital literacy, making them more susceptible to privacy threats [20]. Additionally, many countries lack robust data protection laws, offering fewer safeguards [21]. Furthermore, infrastructure constraints led to a lack of internet connectivity and issues with the power supply, which may compromise the security measures [22].

In Saudi Arabia, privacy perceptions are shaped by strong religious and cultural values [23]. Islam defines societal norms [24], while Bedouin traditions emphasize kinship and extended family ties [25]. This context influences privacy practices, as families often view privacy as essential to protecting honour and reputation[26]. Gendered privacy norms are also significant, with women experiencing heightened concerns due to expectations of modesty, which affect their clothing, online presence, and behaviour.

The collectivist nature of Saudi society reinforces community influence, where personal actions are scrutinized and social conformity is expected. This dynamic blurs the line between public and private, embedding privacy within broader cultural values like modesty and family honour. As scholars note, privacy concepts vary across cultural and geographical contexts [27]. In Saudi Arabia, gendered privacy is especially prominent as women show more online caution due to cultural expectations [28]. Moreover, Saudis express strong digital privacy concerns, avoiding platforms they see as intrusive [29], often using pseudonyms [30].

Islamic teachings also shape privacy views, emphasizing modesty, segregation, and the sanctity of personal and family data [31]. Collectivism further influences behaviour, with societal norms pressuring individuals to align with community expectations to avoid reputational harm [30]. In sum, these studies demonstrate that Saudi Arabia's cultural, religious, and societal factors shape privacy perceptions and practices differently from more individualistic cultures. However, there is limited evidence on how cultural factors influence the concept of privacy, particularly from the perspective of the user.

With the growth of IoT, privacy concerns around smart speakers have intensified [32]. Surveys indicate that users in Saudi Arabia, particularly those in the financial and public sectors, have experienced privacy violations, with 42% and 24% of respondents, respectively, reporting incidents. Additionally, nearly half of the participants demonstrated a lack of awareness regarding Saudi IT criminal laws [33]. These concerns align with global trends, as privacy remains a key issue across regions such as the United States, Europe, and Asia. Studies have shown that individuals place a high value on privacy, particularly in the digital era [34]. Notably, 84% of Saudis consider privacy important, similar to EU findings under GDPR. Furthermore, 29% distrust online shopping [33], reflecting broader global concerns about fraud and data breaches [35, 36]. U.S. and UK users also show mistrust toward government websites' data practices [37] [38].

Cultural norms and gender dynamics in Saudi Arabia may heighten discomfort around sharing personal data, especially compared to other regions where gender gaps are less pronounced [39]. Government overcollection of data also appears more severe in Saudi Arabia, where users report excessive data collection for basic tasks [35]. By contrast, the EU's GDPR enforces data minimization [40]. To date, no research has specifically explored smart speaker privacy concerns in Saudi Arabia. This study addresses that gap by examining the current landscape among Saudi users.

### 2.4 Overview of cultural probes

Human-Computer Interaction (HCI) has recently expanded beyond workplace and efficiency-focused research to explore everyday life, including domestic, urban, play, and entertainment contexts [41]. In conjunction with this shift to new areas of investigation, innovative methods have been developed for interacting with them. Among these, the use of "probes" stands out as one of the most notable and influential approaches [42]. Probes were first developed by Bill Gaver's team during the EU Presence Project, which aimed to engage older adults in community life [43].

Due to the team's limited ability to spend extended time in communities, they created "cultural probes." These probes served as a design-oriented method for obtaining inspirational insights into the communities being studied. The cultural probes enhanced engagement with participants and were introduced after several initial interactions had taken place. Designed as physical kits, these probes included a range of open-ended, thought-provoking, and indirect tasks aimed at fostering early participant involvement in the design process [43].

Prior smart speaker privacy studies rely on interviews or questionnaires, which can miss the subtle, everyday dynamics of privacy management. To address this, we combined semi-structured interviews with cultural probes for uncovering context-specific, personal, and often implicit behaviours over time. This approach is particularly valuable in Saudi Arabia, where privacy practices are shaped by cultural norms, gender roles, and social expectations. Probes offer a flexible and culturally sensitive method to explore how users engage with privacy in their homes.

## 3 METHODS

We conducted a qualitative study using cultural probes and semi-structured interviews with 16 participants from Saudi Arabia, aged 18 to 55. The participants were recruited from two key regions: the capital city, Riyadh, and Jeddah two major urban centres selected for their population density and distinct cultural profiles. The study aimed to address the research question: What privacy-protection behaviours do Saudi users adopt when interacting with smart speakers, and how aware are they of related privacy concerns? Cultural probes were used to elicit rich, contextual insights into participants' everyday interactions with smart speakers, particularly in relation to privacy practices and attitudes. These probes were distributed and completed over a two-week period, after which semi-structured interviews were conducted to explore emerging themes in greater depth. All interview data were transcribed and subjected to thematic analysis [43], allowing us to identify recurring patterns and categorize participant responses. Given the limited research on privacy protection behaviours in non-Western contexts, thematic analysis was chosen for its flexibility and suitability in uncovering nuanced, culturally embedded practices. We conducted a thematic analysis following Braun and Clarke's six-phase framework [44] to identify meaningful patterns in the data. The lead researcher, who is Saudi and shares cultural and linguistic commonalities with participants, played a central role in data collection and analysis. This insider perspective provided valuable contextual sensitivity but also required ongoing reflexivity to critically examine assumptions and interpretations. Ethical approval for the study was obtained from the Human Research Ethics Committee (HREC) of [withheld].

### 3.1 Recruitment

For the purpose of recruiting participants, we used social media sites such as Facebook. In addition, we utilized a faculty member at a Saudi university in order to find participants for the study. A total of sixteen participants were recruited to participate in the study. An overview of the participants can be found in **Table 1.**

**Table 1.** Summary of the participant demographics (N=16).

| Participant | Sex | Age | Type of smart speaker | Length of use | Household Profile (living alone, shared house with friends, family, family with kids) |
|---|---|---|---|---|---|
| **P1** | Male | 19 | Alexa | 1 year | Shared house with family |
| **P2** | Male | 20 | Alexa | 8 months | Shared house with family |
| **P3** | Male | 24 | Alexa | 3 years | Shared house with friends |
| **P4** | Male | 25 | Alexa | 5 years | Shared house with family |
| **P5** | Female | 26 | Alexa | 2 years | Shared house with family |
| **P6** | Male | 29 | Alexa | 4 years | Living alone |
| **P7** | Male | 32 | Google Home | 10 months | Family without kids |
| **P8** | Male | 34 | Alexa | 1 year | Family without kids |
| **P9** | Male | 36 | Alexa | 2 years | Family with kids |
| **P10** | Male | 36 | Google Home | 2 years | Shared house with friends |
| **P11** | Male | 37 | Google Home | 2 years | Shared house with family |
| **P12** | Female | 38 | Alexa | 1 year | Shared house with family |
| **P13** | Male | 41 | Alexa & Google Home | 8 years | Family with kids |
| **P14** | Male | 43 | Alexa | 4 years | Shared house with friends |
| **P15** | Female | 45 | Alexa | 9 months | Family with kids |
| **P16** | Male | 55 | Alexa | 3 years | Living alone |

### 3.2 Culture Probes

As part of our effort to better understand participants' perceptions about and experiences with smart speakers, we conducted a cultural probe study [43]. This method involves participants participating in an open-ended and engaging activity on their own time, allowing them to contribute to our work by telling us how they utilize smart speakers in accordance with the research question. Therefore, the use of this formative research method assists us in building a more individualised understanding of participants' experiences as well as identifying unexpected perceptions, attitudes, and behaviours. In this study, the aim was to determine what privacy protection behaviours participants employ when using smart speakers.

Participants were provided with a probe package designed to be completed over a seven-day period. The package contained a variety of materials aimed at exploring privacy protection behaviours, including the following:

1. **Maps**: Participants were given labelled map templates to annotate with notes such as "Mark the location of your smart speakers" or "Identify areas where you feel your privacy is being compromised." Alternatively, participants could draw their own maps. They were encouraged to describe each room or area in their home, detailing any privacy concerns or considerations related to the use of smart speakers. Participants could include additional notes and clarifications, explaining their choices and the associated privacy implications. Figure 1 provides an example of a map created by one of the participants.

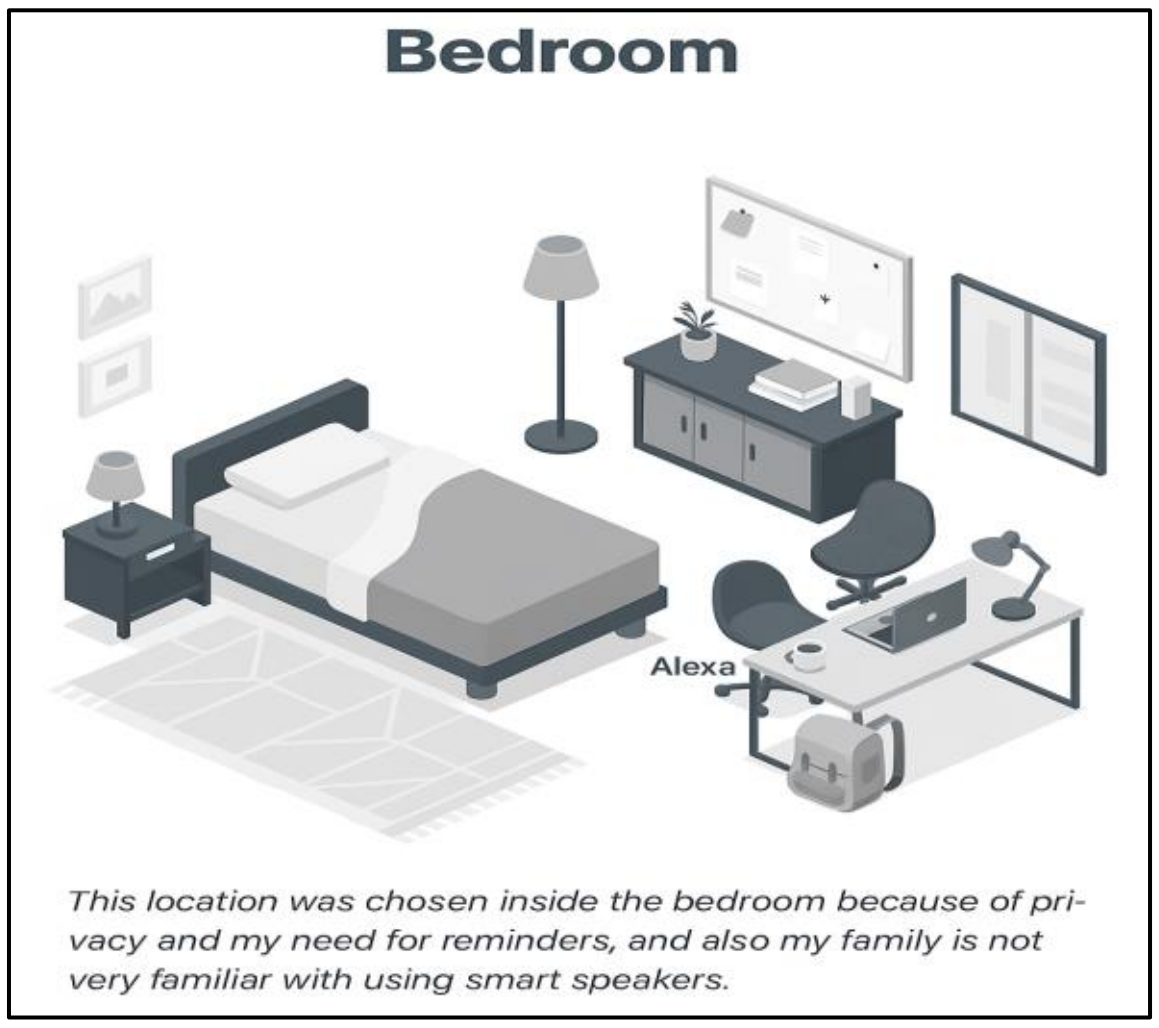

Figure 1: P3 generated map.

2. **Scenario-Based Prompts**: Five hypothetical scenarios were provided for participants to respond to, requiring them to describe their actions and reactions to various privacy-related situations. These scenarios were informed by prior research on privacy behaviours, smart speaker usage, and cultural factors [15] and [45]. They were designed to reflect realistic situations that align with previously identified concerns and contextual factors reported in prior studies. Additionally, participants were asked to document one real-life scenario they had experienced and explain their responses. Each scenario was intended to be completed daily. Scenario 1 explored user reactions to a smart speaker unexpectedly activating and appearing to record a private conversation. Scenario 2 examined responses to receiving product or service recommendations on their phone based on nearby conversations, despite no related online searches. Scenario 3 involved the smart speaker overhearing a discussion about a family gathering and autonomously ordering food and supplies. Scenario 4 asked how users would respond if a third party remotely accessed the device. Scenario 5 focused on a guest accessing personal data (e.g., schedules, contacts, or payments) without consent. Finally, scenario 6 invited users to describe and reflect on a real-life experience. The full list of scenarios is included in (**Appendix A)**.
3. **Privacy Diaries**: Participants were given diary templates to record daily experiences and reflections on privacy while using smart speakers.
4. **Open-Ended Prompts**: Blank sheets of paper were provided to participants with a general instruction to freely express any thoughts, concerns, or ideas they had about privacy in the context of smart speakers. This allowed them to reflect on and share their personal experiences, worries, or suggestions without being guided by specific questions.
5. **Participant Information Sheet**: Participants were provided with detailed information about the study's purpose, data collection methods, and data usage. They were required to read and sign a consent form to confirm their participation.

Figure 2 illustrates examples of the materials included in the probe package. Participants returned their completed probe packages one week after the experiment concluded, with the exception of three participants who required an additional three days to complete the tasks. All collected materials were transcribed, encoded, and converted into PDF files for analysis.

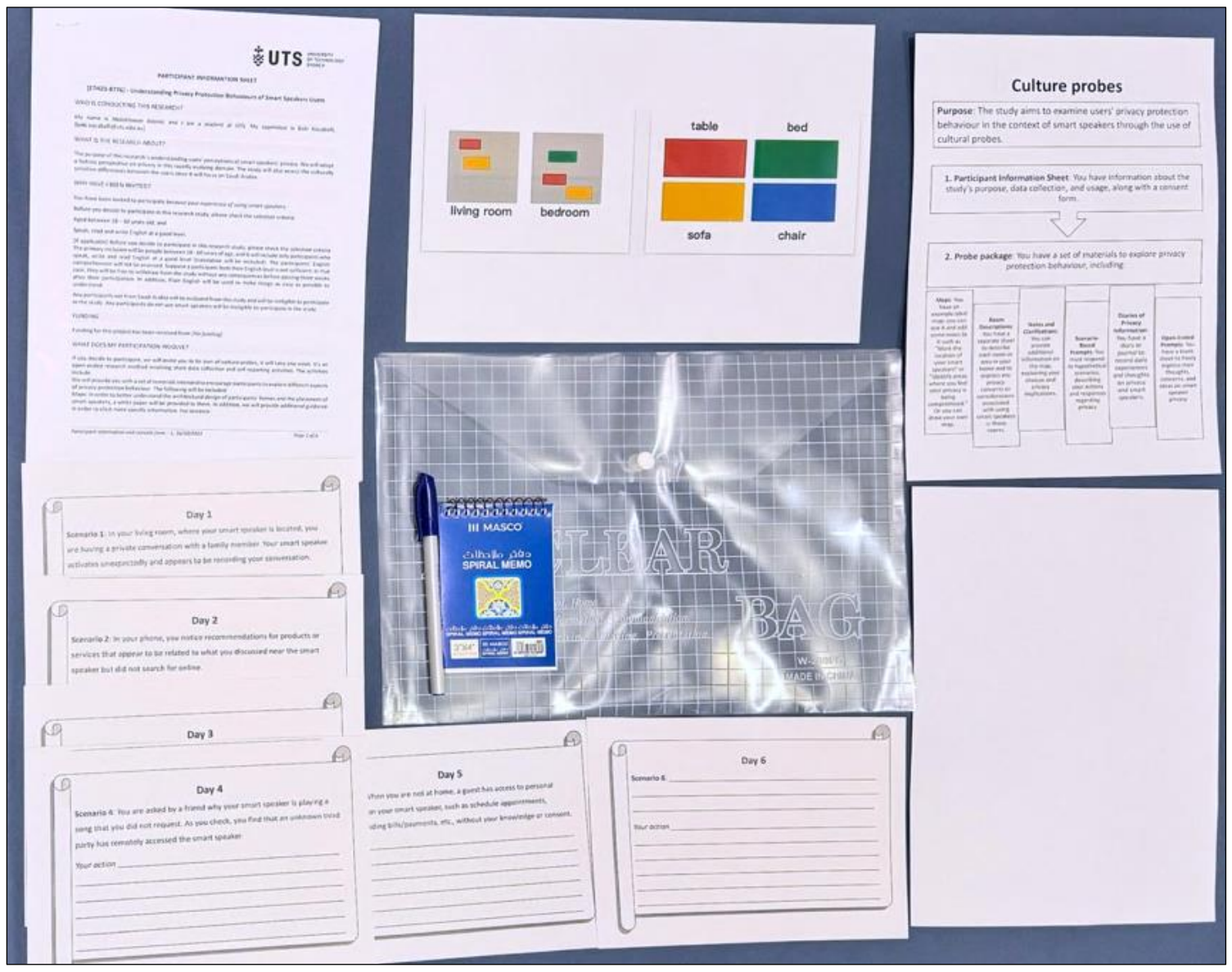

Figure 2: The content of the probe package.

### 3.3 Semi-structured interviews

The cultural probe investigations provided valuable data that informed the development of the semi-structured interviews. Drawing from the analysis of the data gathered during these investigations, alongside a review of relevant empirical literature, an interview guide was created. The interviews addressed a wide range of topics, including the placement of smart speakers, motivations for adopting them, patterns of use over time, privacy concerns, privacy settings, protective behaviours regarding privacy, and tangible privacy. The interview consisted of 27 questions (**Appendix B**).

Between August and September 2024, 16 participants participated in semi-structured interviews online via Zoom. Each interview, facilitated by the first author, had an average duration of 35 minutes. All sessions were digitally recorded and subsequently transcribed for analysis.

### 3.4 Data analysis

To streamline the analysis process and enable seamless collaboration among the authors, we utilized the Miro app [46] to organize and present all materials received from participants in the cultural probe study. This approach facilitated content sharing and enhanced the overall analysis workflow.

We conducted an inductive, reflexive thematic analysis through a systematic and iterative process. The analysis began with familiarization and initial coding of the cultural probe materials, which informed the coding of interview transcripts. The first and second authors, supported by a research assistant, independently reviewed and coded the transcripts, engaging in regular analytic discussions to refine and consolidate codes. A thematic framework was collaboratively constructed by the first and second authors. NVivo software was used to facilitate data coding and management.

During the data collection and analysis phases, we observed that thematic saturation was reached by the time we completed interviews with the sixteenth participant. Thematic saturation refers to the point at which no new themes, insights, or codes emerge from the data, indicating that further data collection is unlikely to provide additional relevant information [47]. At this stage, the responses had become repetitive and consistent, suggesting that the data was sufficiently rich to address our research questions. Additionally, the analysis process was informed by a reflexive approach. As the lead researcher is culturally familiar with the Saudi context, we remained mindful of potential biases when interpreting participants' privacy concerns and behaviours. To enhance credibility and reduce individual interpretation bias, two researchers reviewed and discussed the data collaboratively. These discussions supported theme refinement and helped ensure that the findings were grounded in the participants' perspectives rather than solely shaped by researcher assumptions.

## 4 RESULTS

### 4.1 Major Themes from Cultural Probes

The cultural probe study revealed nuanced insights into user experiences, attitudes, and behaviours surrounding smart speaker use and privacy protection across several themes: distrust and defensive actions, personalization, spatiality, child and family use and.

**Distrust and defensive privacy actions.** Distrust emerged as a dominant response across several scenarios, particularly those involving unexpected device behaviours. In Scenario 1, where the smart speaker activated without a prompt, participants expressed immediate discomfort and took defensive actions: "*I will stop the conversation and check the settings and delete the history*." (P1), "*I'll unplug the device and warn others in the room*." (P7). In many cases, participants chose to turn off, mute, or disconnect the device completely, often stating explicitly that they didn't trust it: "*I faced this scenario before… I paused the conversation till I turned it off*" (P12).

**Personalization: Helpful Yet Intrusive.** The utility of personalized features (Scenario 2) presented a notable tension, frequently embodying a 'privacy paradox' where perceived benefits clashed with feelings of unease. While tailored recommendations were sometimes appreciated for their relevance, they often coexisted with a disquieting sense that the device was overly invasive. As (P3) articulated, "*Sometimes it's helpful, but I feel like it's always listening*." This discomfort prompted some users to adopt circumvention strategies, such as deactivating the assistant during private discussions: "*It happens to us many times... we discussed it and now we turn the assistant off when we have private conversations*."(P11). Furthermore, proactive device behaviours initiated without explicit user consent (Scenario 3), such as the smart speaker independently ordering items, were met with strong negative sentiment. Participants viewed such actions as clear oversteps, expressing intentions to "*cancel the order and remove my credit card*"(P13) and labelling the event a "*privacy violation,*" with some stating they would "*disable shopping features*." These findings highlight a critical need for enhanced user control and transparent consent mechanisms, particularly as smart speakers exhibit greater agency.

**Spatiality, Social Boundaries, and Device Placement.** The physical placement of smart speakers within the domestic environment emerged as a key strategy for negotiating privacy, especially concerning access by third parties or guests (Scenarios 4 and 5). Visual data from the probes illustrated a clear demarcation by participants between 'shared' and 'private' zones within their homes. Many (P3, P5, P10) chose to locate their devices in more secluded areas such as bedrooms or studies, explicitly citing privacy as the rationale: "*I keep it in the bedroom. It's not convenient to turn it on and off every time, but I don't want it on in shared spaces*" (P3). Conversely, speakers positioned in communal areas like living rooms were more frequently associated with anxieties about unintentional eavesdropping or activations.

**Family Use and Control.** Several responses focused on children's interaction with smart speakers. Participants reported use cases such as playing music or asking educational questions: "*My kids asked it to answer questions from school.*" (P9).

However, others expressed concerns about unsupervised access and content filtering: "*Can children have unlimited access? Can it be changed?*" (P13). This ambivalence reflects a dual framing of the device as both an educational aid and a potential source of risk, complicating how families regulate its use.

### 4.2 Major Themes from Semi-structured Interviews

Our semi-structured interviews with smart speaker users in Saudi Arabia revealed five central themes that characterize their experiences: A) Placement of Smart Speakers, B) Usage of Smart Speakers, C) Privacy Concerns, D) Privacy Protection Behaviours, and E) Tangible Privacy Solutions. The following section will provide a more detailed explanation of each theme.

#### *4.2.1 Strategic Placement of Smart Speakers: Negotiating Privacy, Accessibility, and Control*

Participants' decisions regarding smart speaker placement were driven by a confluence of practical, personal, and privacy considerations, reflecting an ongoing negotiation between privacy, accessibility, and perceived control.

Many participants positioned smart speakers in central household locations like living rooms to maximize accessibility and shared use. P6 placed their device "*in the living room, close to the TV, because that's the central location and where I spend most of my time,*" echoing P16's sentiment of needing it "*close by... whenever we need something.*" This contrasted with a significant number of users who opted for private spaces, predominantly bedrooms. This choice was driven by desires for personalized use, such as P9's need to "*make more specific tasks like control some other devices that involve the bedroom,*" and explicit privacy concerns. P1 clearly stated, "*it's in my room because of the privacy. For my family, it is not known about this device.*" P13's relocation of the device from the living room to their bedroom "*because I have a roommate*" further underscores the use of placement to control social exposure.

Placement choices also reflected household compositions and evolving needs. For instance, P9 strategically placed a second device "*in the kids' rooms... to integrate them with the device to lead them to the future.*". Multi-device households (P5: "*We have three Alexa devices... One in the living room, one in my sister's room, and one in my room*") demonstrated tailored placement according to individual family members' spaces. The adaptability of placement was highlighted by P12, who lives alone and moves the device between rooms: "*When I sit in the lounge room, I put it in the lounge room... But when I go to the bedroom, I take it with me.*"

#### *4.2.2 Core Usage of Patterns*

Participants reported using various functionalities and tasks, including home automation, information retrieval, entertainment, and routine management. Participants commonly used smart speakers to control home appliances. P2 used it to operate the AC and TV. P9 set timers for the AC and lights in the bedroom. P8 highlighted its usefulness in hot climates, using it to pre-cool the room and control devices like TVs. Also, smart speakers provided quick access to information. P2 used it for weather updates and general questions. P16 relied on it for fast information sources, and P14 used it for small queries like time and weather. Many participants used smart speakers for entertainment. P4 listened to music, P13 watched sports matches via commands, and P3 played bedtime stories for their child. Moreover, smart speakers assisted with reminders and schedules. P9 set reminders for diabetes medication, P7 scheduled appointments, and P8 used reminders for appointments. Some faced difficulties with voice control. P9 struggled with pronunciation, and P14 noted occasional inaccuracy when managing multiple devices like lights. Participants also used smart speakers for tailored tasks. P12 quickly found shopping information, while P6 set routines for bedtime, like turning off lights and locking doors. Participants interacting with smart speakers exhibited several key usage patterns. Device control and home automation were common uses. Information retrieval was another frequent behaviour. Many participants used smart speakers for entertainment. Reminders and scheduling were also significant. Several participants reported pronunciation issues and inaccuracy when controlling multiple devices. Moreover, a group of users customized their experience by automating bedtime routines and searching for shopping information. These patterns illustrate the multifunctionality of smart speakers and the need for improved voice recognition and personalization capabilities.

#### *4.2.3 Privacy Concerns*

Privacy was a pervasive concern, expressed in both general unease and specific anxieties. Participants were wary of unintended activation, with several noting instances where the device appeared to listen without prompt. P2 remarked, "*It activates unexpectedly,*" echoing broader worries about passive listening. Concerns over data handling were prevalent, particularly around data being collected, stored, or accessed without transparency or consent. P14 questioned, "*If you collect my data or logs, at least let me know.*" Distrust extended to built-in privacy controls. Many doubted whether features like mute functions were effective, choosing to unplug the device altogether during sensitive conversations. P15 explained, "*I didn't trust the mute button… I just unplug the speaker.*" Participants were especially cautious about storing sensitive data such as credit card details, with P8 expressing fear that this could lead to fraud.

Context played a key role in shaping participants' concerns. Bedrooms, for example, were often considered inappropriate for device use due to their perceived intimacy. Cultural factors also influenced perceptions, particularly when guests were present. P8 recalled a situation where a female visitor felt uncomfortable knowing a camera-enabled device was nearby, reflecting broader sensitivities around gender norms and surveillance. Notably, some participants exhibited a privacy paradox: downplaying concerns in one moment while revealing deep-seated unease later. P13 initially claimed, "*I don't mind,*" but later acknowledged fears about password security and the risks of data leakage. This disconnect often reflected a tension between the desire for convenience and unease with data practices.

#### *4.2.4 Privacy Protection Behaviours: Practical Workarounds and Awareness Gaps*

Participants described a spectrum of strategies ranging from quick physical gestures to more deliberate configuration changes to keep smart-speaker surveillance under control. Across accounts we observed six recurring practices, each revealing different degrees of effort, perceived efficacy, and underlying awareness.

**(i) Toggling the microphone.** Using the mute button or turning off the device was a common method for protecting privacy. P2 stated, "*I put it in mute mode*.". P12 elaborated, "*I make sure always to put it in mute mode if I don't need to use it*.". P4 shared, "*If the smart speaker activates unexpectedly... I just use the mute button*.". P14 explained, "*If I want to hide something, I just will press the mute button. Or I will unplug it from the electricity to make sure that it will not work*."

**(ii) Pulling the plug.** Many participants chose to unplug or completely power down their devices during sensitive times. P6 mentioned, "I turned it off and disconnected it from the Internet.". P15 added, "*I just unplug the speaker*.". P7 explained, "*I know I said that many times... but this is my first solution, even sometimes I ignore this mute button, and just plug it off*.".

**(iii) Behavioural Caution.** The participants exhibited caution when interacting with smart speakers, avoiding sensitive topics or limiting the device's capabilities. P12 noted, "*The measure to protect myself is to always have to be careful about what information should be asked and what information should be given to the smart speaker*.". P9 remarked, "*Most likely I try not to do specific tasks that are a concern for my privacy, so I try to be aware of that*.". P16 stated, "*I try not to expose a lot of private or sensitive information while I'm using the smart speaker*.".

**(iv) Monitoring, Adjusting and Engagement with privacy settings.** Roughly half of the sample reported checking logs or tweaking settings (P10, P11, P6). The other half admitted ignorance or frustration: "*it's very difficult. Because there are too many choices, and you have to search for where is the best one for you*." (P13). This split indicates a usability gap in vendor dashboards: controls exist, yet remain effectively hidden for many.

**(v) Use of Passwords and Authentication.** Several participants enhanced the security of their accounts by using passwords and other authentication methods. P8 stated, "*I used a password to protect my private personal information, so it will help me to keep my data safe*.". P16 added, "*Using my password to protect my private information and personal information makes me feel more comfortable and secure*.".

**(vi) Limiting Access and Placement.** Finally, participants limited who could speak to the device by relocating it: "*I put it in the bedroom to make it limited... so if someone comes from outside, it's difficult to use it.*" (P9). P1 mentioned, "*When someone comes to my room, I shift it off to protect my privacy*". Ultimately, inconsistent understanding and use of privacy features pointed to a usability gap in privacy management.

Taken together, these practices expose two intertwined gaps: a design gap (complex or obscure settings) and an awareness gap (incomplete mental models of how data are captured). Closing either requires not only better interface affordances but also scaffolding that turns ad-hoc workarounds into confident, informed choices.

#### *4.2.5 Tangible Privacy Solutions*

When presented with the concept of tangible privacy solutions such as physical objects designed to offer users more direct and visible control over device data collection (e.g., a wearable ring or necklace to mute/block data collection), participant reactions were diverse, ranging from strong enthusiasm to cautious scepticism, often coloured by practical and cultural considerations. Several participants expressed strong interest, valuing the potential convenience and simplicity. P1 saw it as a way to "*make my life easier since it's easier just to push or touch the ring.*" P3 believed "*some people will like the idea and have the chance to easily mute the device without reaching out.*" Others, while open, expressed a need for trial or additional functionality. P12 felt, "*It could be a valuable solution, but I have to try it before,*" while P14 considered its appeal greater "*If it has more features*" beyond just privacy, also raising concerns about the physical safety of wearable tech.

Conversely, some participants were sceptical about the effectiveness or necessity of such solutions. While P6 stated *"I'm not sure how effective it would be"*), P5 suggested *"the current one [mute button] is better."* Cultural context also influenced participants' opinions. P1 commented, "Maybe that's good as a solution." However, when asked about personal use, P2 explained, "*From a cultural perspective, maybe I will not use it as these things more suitable for women*.". He was referring specifically to devices designed as rings or necklaces, which he perceived as feminine accessories.

## 5 DISCUSSIONS

### 5.1 Understanding Privacy Behaviours: Insights from Prior Research.

The findings of the study revealed several key strategies, such as the use of mute buttons, power disconnection, cautious communication, and engagement with privacy settings. This behaviour emphasizes the user-driven nature of privacy, necessitated by a lack of trust in the functionality of smart speakers. the findings of this study are aligned with the results of previous research studies [48, 49] [50-52]. Moreover, participants lack confidence in the assurances provided by device manufacturers regarding privacy because they frequently use the power-off and mute functions. The findings of this study are also consistent with existing research indicating that users resort to physical disconnects when they feel that digital safeguards are inadequate [53, 54]. These findings highlight the need for device designs that promote greater transparency and trust among users. Several promising approaches have emerged in recent research to enhance transparency and build user trust in smart speakers and artificial intelligence assistants. One such innovation is the Candid Mic, a battery-free wireless microphone powered by energy harvested from intentional user interactions. Unlike traditional mute buttons, this design offers a perceptible indication that the microphone is active, thereby providing users with greater assurance regarding when audio is being recorded[19]. Privacy-aware response generation is another promising idea. Researchers have proposed integrating preprocessing techniques with Natural Language Understanding models in the context of English Language-Based Virtual Assistants (ELB-VAs). The goal of this approach is to ensure that the interaction is both informative and privacy-conscious through the incorporation of privacy-aware response generation [55].

Users of smart speakers employ behavioural caution, such as avoiding sensitive topics near them, in order to minimize the risks associated with their use. While such strategies are effective, they place the burden of responsibility on the user and highlight a gap in the functionality of the device. Researchers and designers of smart speakers should explore ways of reducing this cognitive load through the integration of privacy-friendly features [56]. The act of monitoring and adjusting privacy settings was widely reported as a common practice [57]. However, many participants expressed significant frustration with the overwhelming complexity of these settings. This aligns with the findings presented in [56], underscoring the critical need for designing user interfaces that prioritize simplicity and accessibility in privacy controls. By implementing streamlined settings and offering proactive guidance, users can confidently and effortlessly assert control over their data security.

For instance, PriviFy is a tangible interface prototype that employs physical elements such as knobs, buttons, lights, and notifications, enabling users to configure their data privacy preferences intuitively [58]. By leveraging tactile and visual controls, this approach reduces the complexity typically associated with privacy configurations. PriviFy was evaluated for its effectiveness in simplifying IoT privacy configuration. As compared to the traditional digital interface (Alexa Application), PriviFy achieved significantly higher usability scores, with a SUS score of 82.8, indicating an excellent level of usability (p .001). According to the User Experience Questionnaire (UEQ), it outperformed the digital interface on all dimensions, including attractiveness, efficiency, and dependability. PriviFy enabled participants to locate privacy settings more quickly and independently, and qualitative feedback revealed that users preferred its physical controls, clarity, and ease of use. Based on these findings, PriviFy has the potential to improve user engagement, trust, and privacy control through tangible interactions. Similarly, the Privacy Assistant offers a conversational interface that uses natural language interaction to guide users through the privacy settings of smart speakers [59]. The study found this design is particularly effective in making privacy controls more user-friendly, especially for older adults, by providing clear explanations and step-by-step assistance.

The complexity of privacy settings can be attributed to several key factors. One significant issue is the presence of multiple menus and unclear labelling, which often deter users from engaging with these controls [58]. Additionally, a misalignment of incentives between users and device manufacturers contributes to this challenge, as manufacturers may prioritize the collection of personal data for their benefit, creating a conflict of interest that undermines user-centred privacy design [19]. Another method, proactive guidance, involves systems taking the initiative to inform and support users in managing their privacy. For example, voice-based Experience Sampling Method (ESM) applications actively engage users to document interactions and identify potential errors [60]. Similarly, smart speakers can provide timely reminders and facilitate conversations related to chronic disease management [61]. However, implementing proactive guidance comes with challenges, such as maintaining a balance between proactive assistance and user autonomy, avoiding user annoyance, and ensuring that these features do not introduce new privacy risks.

Several promising strategies for enhancing privacy controls in smart speakers have been identified. One key approach is the implementation of customizable privacy settings, which allow users to tailor both their privacy preferences and content [62]. Studies indicate that this level of customization fosters greater trust among users, suggesting opportunities for further development of personalized and user-friendly privacy controls. Another noteworthy direction involves conversational interfaces. Voice-based privacy assistants, such as those designed to support older adults, demonstrate significant potential in improving the accessibility of privacy settings by leveraging natural language interactions [59].

To drive progress in this field, future research should prioritize the development and evaluation of more intuitive, tangible, and conversational interfaces for privacy management. Additionally, researchers should explore methods for delivering proactive guidance that respects user autonomy and avoids introducing new privacy vulnerabilities. Addressing the unique challenges posed by multi-user environments and cultural differences in privacy expectations is equally crucial. Furthermore, examining the long-term effects of simplified privacy controls on user behaviour and trust in smart speakers will provide valuable insights for designing more effective and user-centred solutions.

The privacy paradox occurs when individuals express indifference to privacy concerns, but manifest underlying apprehensions through their actions or deeper discussions. According to our study, the results are in agreement with previous research. For instance, Seniors who use smart speakers and mobile health apps dismissed privacy risks due to resignation (e.g., claiming that they would not be affected by a leak of health information later on in their lives) or lack of technical expertise [63]. Further investigation, however, revealed concerns regarding the intrusion of personal information and physical security. Additionally, prior research has demonstrated similar trade-offs between privacy concerns and perceived benefits for smart speakers and other IoT devices. In several studies, users have stated that they are often willing to compromise privacy for the convenience and functionality offered by these technologies, while maintaining reservations regarding certain aspects of their use. For example, in a study on smart home IoT device management, participants noted that privacy was more important to them than convenience, but they continued to utilize these devices. The use of IoT technology implies an acceptance of some privacy risks in exchange for the benefits it provides [64]. Furthermore, the study revealed that different levels of importance placed on privacy control and convenience resulted in varying preferences for privacy control and convenience[64]. Similarly, research has shown that privacy concerns can vary depending on the setting and context of use. According to a study on privacy-preserving interventions for smart speakers, users had to make complex trade-offs between utility, privacy, and usability [65]. Consequently, it may be that people are more willing to use smart speakers in certain areas of their homes, while being more hesitant to use them in private places.

### 5.2 The influence of culture on privacy aspects

Several cultural dimensions have been identified as influencing user practices, including device placement, usage, and privacy management strategies as well as the responses to tangible privacy solutions. In the context of Saudi households, smart speakers were typically placed in communal areas rather than private spaces. This behaviour aligns with cultural norms emphasizing modesty and the protection of personal boundaries [66]. While similar placement decisions may also occur in other countries due to general privacy concerns, participants in this study explicitly linked their choices to local cultural and religious values, suggesting that these norms played a distinct role in shaping their practices. This sentiment aligns with Islamic principles emphasizing the protection of personal and familial privacy. Participants often placed smart speakers in shared spaces, such as living rooms, to enable controlled use while preserving cultural sensibilities. However, this practice is not uniform. While some saw shared areas as less intrusive, others acknowledged that these spaces could expose private conversations to guests, raising additional concerns. From this perspective, placing the device in a bedroom might paradoxically offer more privacy by limiting exposure to non-family members. This reflects the multidimensional nature of privacy, which is not only about physical space but also about contextual boundaries and information flow, as described in Nissenbaum's theory of contextual integrity [11]. Furthermore, Altman's privacy regulation theory [67] suggests that individuals dynamically manage privacy by adjusting access to themselves based on social and environmental contexts which helps explain why some participants were willing to relocate devices depending on who was present or affected. These examples illustrate that privacy in this context is not static, but actively negotiated according to cultural, situational, and relational factors.

Concerns rooted in cultural and societal norms were particularly evident when technology intersected with traditional social interactions. P8's account of a female guest's discomfort with a camera-equipped smart speaker during a visit, in a cultural context valuing modesty and gendered privacy, illustrates how smart technologies can unintentionally clash with cultural expectations around hospitality and privacy, especially for women [68]. Apprehensions about data misuse were also frequently linked to a broader skepticism about the alignment of opaque technological practices with deeply held cultural values, emphasizing a need for transparency within culturally sensitive frameworks [69].

In response, Saudi users adopted proactive strategies like unplugging devices or using physical controls during social gatherings, reflecting a preference for tangible disconnection as a more reliable means of ensuring privacy within the household [53]. Reactions to proposed tangible privacy solutions (e.g., privacy-control rings or necklaces) were also filtered through a cultural lens. While some were enthusiastic, others expressed reservations based on gendered perceptions of accessories, highlighting that for solutions to be broadly accepted, they must be not only functional but also culturally sensitive and inclusive [70]. This aligns with broader research on the impact of cultural perceptions of gender in wearable technology design [71] and comparative studies demonstrating that technology acceptance and preferences for privacy features can differ significantly across cultures [72].

### 5.3 Design Implications

The aspirations voiced by our participants for more innovative and intuitive privacy controls such as physical toggles integrated into everyday objects, distinct privacy modes, intelligent settings, and clear visual alerts signal a clear demand for advancements in smart device design that are context-aware and user-centric. Therefore, several emerging design directions resonate with the needs identified in our study:

**Tangible and Perceptible Controls**: As suggested by the positive reception to concepts like Candid Mic [19] and PriviFy,[58] and the cultural preference for physical disconnection, tangible interfaces offer a promising avenue. Future designs should explore easily accessible physical mechanisms that provide unambiguous feedback about the device's listening state and data collection practices.

**Simplified and Transparent Settings:** The frustration with complex privacy menus calls for radically simplified interfaces. Conversational agents like the Privacy Assistant [59] offer one path, but visual interfaces must also be rethought to prioritize clarity, offer proactive explanations, and minimize cognitive load.

**Culturally Sensitive Design:** The influence of cultural values on device placement, use, and perceptions of privacy solutions cannot be overstated. Design processes must incorporate methods to understand and accommodate these values, considering aspects like gender norms, hospitality practices, and communal versus private space dynamics [70-72].

**Context-Aware Privacy:** Solutions that leverage contextual cues, such as gaze direction or voice volume for selective activation [73], or head orientation sensing like HeadTalk [74], could address concerns about accidental activations and constant listening, thereby aligning better with users' intentions and reducing the need for constant manual intervention.

Lightweight, privacy-preserving authentication protocols, perhaps adapted from sectors like IoT healthcare [75], could also bolster trust. This may involve co-design practices with target user groups and offering customizable features that allow users to align device behaviour with their cultural comfort levels. We summarise culturally grounded implications and feature suggestions in Table 2.

Table 2. Saudi-Specific Insights, Design Implications and Illustrative Features

| Saudi-Specific Insight | Design Implication | Illustrative Feature |
|---|---|---|
| **Concern about guests** | “Guest Mode” control | Auto-mute when unfamiliar voices detected |
| **Gendered presence norms** | Gender-aware visual feedback | LED indicators adaptive to user profile or context |
| **Room use varies by time/event** | Mobile privacy zones | Context-based rules (e.g., for prayer times) |
| **Hierarchical, multi-user homes** | Adaptive consent & visibility | Household dashboard with per-user permissions |
| **Rejection of feminised wearables** | Culturally sensitive tangible controls | Privacy toggles embedded in domestic artefacts |

## 5 LIMITATIONS

The interviews were conducted in English, despite Arabic being the primary language of Saudi participants. This may have created a language barrier that limited some participants' ability to fully articulate their challenges. Furthermore, the findings reflect the views of 16 participants, most of whom were male, and may not fully capture the diversity of experiences and attitudes within the broader Saudi community. Future research should aim to broaden the participant base to include more diverse demographics within the region and conduct comparative studies across different cultures to further delineate universal versus culturally specific privacy needs and behaviours. Additionally, the study relied on self-reported data and reactions to scenarios; observational studies of in-situ smart speaker use could provide deeper insights into actual practices versus stated intentions.

## 6 CONCLUSION

Our study provided an in-depth understanding of the privacy protection behaviour of smart speaker users in Saudi Arabia, identifying the impact of some cultural and social factors on users’ smart speaker usage behaviour. There was an inherent tension between the convenience offered by smart speakers and the privacy concerns they raise. Various strategies were employed by users to mitigate privacy risks, including physical disconnection, cautious communication, and engagement with device settings, but many of these measures reflect a lack of trust in the devices' privacy assurances. The context of a user's culture appeared to have influenced their behaviour, such as how smart speakers are placed to balance accessibility and privacy concerns. The adoption of tangible privacy solutions requires alignment with cultural norms and user preferences. These results have implications for the design of privacy-aware smart speakers. Transparency, intuitive privacy settings, and culturally sensitive features should be prioritized by designers. Addressing these factors is critical for fostering the responsible adoption of smart technologies in the Saudi and broader Middle Eastern contexts. Future research should explore privacy behaviours across diverse cultural settings to develop universally applicable, yet locally tailored solutions.

**Acknowledgement**

During the preparation of this work the author(s) used ChatGPT 4o in order to improve readability and language of the final draft. After using this tool/service, the author(s) reviewed and edited the content as needed and take(s) full responsibility for the content of the published article.

**Appendix A**

## Day 1

**Scenario 1**: In your living room, where your smart speaker is located, you are having a private conversation with a family member. Your smart speaker activates unexpectedly and appears to be recording your conversation.

*Your action* ________________________________________________________

______________________________________________________________________

______________________________________________________________________

--------------------------------------------------------------------------------------------------------------------------------

## Day 2

**Scenario 2**: In your phone, you notice recommendations for products or services that appear to be related to what you discussed near the smart speaker but did not search for online.

*Your action* ________________________________________________________

______________________________________________________________________

______________________________________________________________________

## Day 3

**Scenario 3:** When your smart speaker overhears a conversation about an upcoming family gathering, it automatically orders supplies or food for it without your explicit direction.

*Your action* ____________________________________________________

______________________________________________________________

______________________________________________________________

---

## Day 4

**Scenario 4**: You are asked by a friend why your smart speaker is playing a song that you did not request. As you check, you find that an unknown third party has remotely accessed the smart speaker.

*Your action* ____________________________________________________

______________________________________________________________

______________________________________________________________

# Day 5

**Scenario 5**: When you are not at home, a guest has access to personal information on your smart speaker, such as schedule appointments, contacts, pending bills/payments, etc., without your knowledge or consent.

*Your action* ________________________________________________

________________________________________________________

________________________________________________________

------------------------------------------------------------------------------------------------------------------------

# Day 6

**Scenario 6**: ________________________________________________

________________________________________________________

______________________________________________________

*Your action* ________________________________________________

________________________________________________________

________________________________________________________

**Appendix B**

Q1: For a starting point, how was it set up? Is there a reason why you placed it there? Have you considered any other locations? Could you please tell me what other members of the household thought about placing it where it is presently located?

Q2: "what was consideration to purchase the device? Was there any consideration for other people (for example: children, roommates, significant other) in your decision-making process? "

Q3: How would you describe your experience so far with the smart speaker?

Q4:"Do you use your smart speaker to control any other smart home devices? If so, how? "

Q5: "Have you installed any 'apps' (Google Home) or 'skills' (Amazon) on your device? "

Q6: "Do you use your smart speaker to automate any tasks? "

Q7: What is the number of people living in your household? Are the smart speakers accessible to everyone and can anyone use them?

Q8: When guests come over, do they use it? How do they feel about it?

Q9: "Are you discussing it with them (or informing guests that there is a smart speaker in your home)? "

Q10: "If you compare your current use of the smart speaker with when you first began using it, what differences do you see? "

Q11: "How do you use your smart speaker versus your phone or computer/laptop?"

Q12: "What level of comfort do you have with asking your smart speaker to perform your tasks? "

Q13: "While using a smart speaker, do you have any privacy concerns? "

Q14: Has there ever been a time when you forgot that the device was on? If this has happened to you recently, please tell us about it.

Q15: Have you experienced any discomfort when using the device? If this has happened to you recently, please describe the event. Do you have any problem with mute button? where do you think should put it?

Q16: Have you taken any steps to alleviate your discomfort? Could you please describe what you did? Are you satisfied with the solution that has been implemented to address your discomfort?

Q17: When privacy is mentioned as a concern, we will ask the participant the following question. Can you provide me with more information about that?

Q18: When reviewing your smart speaker's history of commands, is there anything that you would like to discuss where you had privacy concerns?

Q19: When using smart speaker such as Google home or Amazon Alexa, is there anything you are concerned about regarding the privacy of your data?

Q20: Are you concerned about the privacy of your data when it is being handled by third party applications?

Q21: In order to ensure your privacy, what measures do you take to protect yourself? What do you think if we use tangible privacy solution such as ring or necklace?

Q22: "Do you ever look at the audio logs collected by your smart speaker? How would you prevent your smart speaker from listening to your conversation? "

Q23: If the smart speaker activates unexpectedly, what action will you take? And there will be a follow question depending on the participants answer. (scenario 1)

Q 24: If the smart speaker provides you with a recommendation for service or product, will you consider it as breach for your privacy or not and why? (scenario 2)

Q25: Do you use the purchase features in the smart speaker, if yes why and if not why? (scenario 3)

Q26: Do you use third party app on your smart speaker, if yes why and if not why? (scenario 4)

Q27: Do you save personal information like your appointments or credit card on the smart speaker, if yes why and if not why? (scenario 5)